\documentclass[11pt]{article}
\usepackage{moriond,epsfig}
\usepackage{color}
\def\Journal#1#2#3#4{{#1} {\bf #2}, #3 (#4)}

\def\NIMA{{\em Nucl. Instrum. Methods} A}
\def\NPB{{\em Nucl. Phys.} B}
\def\PLB{{\em Phys. Lett.}  B}

\def\PRC{{\em Phys. Rev.} C}

\def\PRD{{\em Phys. Rev.} D}
\def\JHEP{{\em J. High Energy Phys.}}

\def\be{\begin{equation}}
\def\ee{\end{equation}}
\def\bea{\begin{eqnarray}}
\def\eea{\end{eqnarray}}

\def\ifm#1{\relax\ifmmode#1\else$#1$\fi}

\def\f{\ifm{\phi}}

\def\x{\ifm{\times}}

\def\pt#1,#2,{\ifm{#1\x10^{#2}}}   

\renewcommand{\to}{\ensuremath{\rightarrow}}

\newcommand{\ep}{\mbox{$e^+$}}
\newcommand{\el}{\mbox{$e^-$}}
\newcommand{\ks}{\mbox{$K_S$}}
\newcommand{\kl}{\mbox{$K_L$}}

\newcommand{\pip}{\mbox{$\pi^+$}}
\newcommand{\pim}{\mbox{$\pi^-$}}
\newcommand{\pio}{\mbox{$\pi^{0}$}}
\newcommand{\fo}{\ensuremath{f_0}}

\newcommand{\dafne}{\mbox{DA$\Phi$NE}}

\newcommand{\kk}{\mbox{$K\bar{K}$}}

\newcommand{\lint}{\mbox{$L_{\rm int}$}}
\newcommand{\wpo}{\mbox{$\omega\pio$}}
\newcommand{\wpc}{\mbox{$\pip\pim\pio\pio$}}
\newcommand{\wpn}{\mbox{$\pio\pio\gamma$}}
\newcommand{\eeto}{\mbox{$\ep\el\to\,\,$}}
\newcommand{\sqrts}{\mbox{$\sqrt{s}$}}

\newcommand{\ket}[1]{\mbox{$\left| #1 \rangle \right. $}}

\newcommand{\etappee}   {\eta~\rightarrow~\pi^+~\pi^-~e^+~e^- }

\begin{document}
%
\vspace*{4cm}
\title{Recent results from KLOE}

\author{ A. De Santis for the KLOE collaboration~\footnote{
\scriptsize
F.~Ambrosino, A.~Antonelli, M.~Antonelli, F.~Archilli, C.~Bacci, 
P.~Beltrame, G.~Bencivenni, S.~Bertolucci, C.~Bini, C.~Bloise, 
S.~Bocchetta, F.~Bossi, P.~Branchini, R.~Caloi, P.~Campana, G.~Capon, 
T.~Capussela, F.~Ceradini, F.~Cesario, S.~Chi, G.~Chiefari, 
P.~Ciambrone, F.~Crucianelli, E.~De~Lucia, A.~De~Santis, P.~De~Simone, 
G.~De~Zorzi, A.~Denig, A.~Di~Domenico, C.~Di~Donato, B.~Di~Micco, 
A.~Doria, M.~Dreucci, G.~Felici, A.~Ferrari, M.~L.~Ferrer, S.~Fiore, 
C.~Forti, P.~Franzini, C.~Gatti, P.~Gauzzi, S.~Giovannella, E.~Gorini, 
E.~Graziani, W.~Kluge, V.~Kulikov, F.~Lacava, G.~Lanfranchi, 
J.~Lee-Franzini, D.~Leone, M.~Martini, P.~Massarotti, W.~Mei, S.~Meola, 
S.~Miscetti, M.~Moulson, S.~M\"uller, F.~Murtas, M.~Napolitano, 
F.~Nguyen, M.~Palutan, E.~Pasqualucci, A.~Passeri, V.~Patera, 
F.~Perfetto, M.~Primavera, P.~Santangelo, G.~Saracino, B.~Sciascia, 
A.~Sciubba, A.~Sibidanov, T.~Spadaro, M.~Testa, L.~Tortora, P.~Valente, 
G.~Venanzoni, R.Versaci, G.~Xu}} 

\address{Dipartimento di Fisica, Universit\'a ``Sapienza'', 
ROMA \& sez. INFN ROMA }

\vspace{-3.0cm}
\maketitle
\abstracts{
We report the newest results from the KLOE experiment on hadronic physics,
such as the parameters of scalars \fo\ and $a_0$, the $\eta$ meson mass 
measurements and dynamics, the first observation of the $\etappee$ rare 
decay, and study of \ep\el\to\wpo cross section around the \f\ resonance.}

\vspace{-0.5cm}
\section{The KLOE experiment}\label{sec:kloe}
\vspace{-0.2cm}

The KLOE experiment~\cite{KLOE} runs at the Frascati 
$\phi$ factory \dafne,
a high luminosity \ep\el\ collider working 
at \sqrts\ $\sim 1020$ MeV, corresponding to the  \f\ 
meson mass. 

\noindent The KLOE detector consists of a large cylindrical drift 
chamber~\cite{DHC}, surrounded by a sampling lead-scintillating fiber 
electromagnetic calorimeter~\cite{EMC}. 
Both detectors operate inside a uniform magnetic field of 
$\sim 0.5$ T provided by a superconducting coil. 
In the whole data taking ($2001 - 2006$) KLOE has collected 
an integrated luminosity of 2.5 fb$^{-1}$ 
corresponding to about 8 billions of \f\ produced, plus 
some \emph{off-peak} data (200 pb$^{-1}$ at 1 GeV). 
The KLOE trigger system~\cite{TRG} is highly efficient on most 
of the \f\ decay. 

\vspace{-0.5cm}
\section{Scalar physics}
\vspace{-0.2cm}

At KLOE the scalar mesons are produced through \f\to$S\gamma$
radiative decays . This allows to observe
the \fo, $\sigma$ and $a_0$ members of the light scalar
mesons multiplet. 

KLOE has already published measurements related 
to decays \fo\to$\pi\pi$, for neutral~\cite{f0p0p0g} and 
charged~\cite{f0pppm} final states. 
Different techniques has been used to analyze the two channels: 
fit to the Dalitz plot density and fit to the di-pion invariant mass,
respectively. In both cases two different phenomenological models
have been used, Kaon-Loop~\cite{Achasov} (KL) and 
``no structure''~\cite{Maiani} (NS).
Agreement between the \fo\ parameters extracted from the two 
channels in the cited analyses are modest. 
A much better agreement between the two channels is 
found using an improved version of the KL model. The update preliminary
for the two channels are shown in Tab.\ref{tab:f0}. 
Currently, we are performing a combined fit of the two spectra.

\begin{table}[h]
  \caption{Updated results for \fo\ parameters.}
  \label{tab:f0}
  \begin{center}
    \begin{tabular}{|l|c|cc|r|}
      \hline
      Channel & Mass [MeV] & $g_{\fo K K}$ [GeV$^2$] & 
      $g_{\fo\pi\pi}$ [GeV$^2$] & R = $(g_{\fo K K}/g_{\fo\pi\pi})^2$ \\
      \hline
      $\fo\to\pip\pim$ & 983.7         & 4.74          & -2.2          & 4.6 \\
      $\fo\to\pio\pio$ & 984.7$\pm$2.1 & 3.97$\pm$0.46 & -1.82$\pm$0.2 & 4.76$\pm$ 0.78 \\
      \hline
    \end{tabular}
  \end{center}
\end{table}
%
%

The $a_0$ meson has been studied~\cite{a0_arch} using the dominant decay
$a_0\to\eta\pio$. Two different $\eta$ decay modes have 
been selected: $\eta\to\gamma\gamma$ and $\eta\to\pip\pim\pio$. The two 
samples are independent and have very different background 
contaminations. Since the interfering $\phi\to\rho\pi^0\to\eta\gamma\pi^0$ 
background is small, it is possible to extract the branching fraction
(BR) directly from event counting after the residual background subtraction:
\be
BR(\f\to a_0\gamma)_{\gamma\gamma} = (6.9\pm0.1\pm0.2)\times10^{-5} 
\qquad 
 BR(\f\to a_0\gamma)_{\pi^+\pi^-\pi^0} = (7.2\pm0.2\pm0.2)\times10^{-5}
\ee
\noindent 
In order to determine the $a_0$ relevant parameters the $\eta\pio$ invariant
mass spectrum, after background subtraction, are being fitted with KL 
and NS parametrisation not reported here.

%
%

Using 1.4 fb$^{-1}$ of the KLOE data, a search~\cite{k0k0g_arch} for 
the decay \f\to\kk$\gamma$ has been performed. In this decay the 
\kk\ pair is produced with positive charge conjugation (\emph{e.g.}
$\ket{i}\propto \ket{\ks \ks} + \ket{\kl \kl}$) and a limited phase 
space due to the small mass difference 
between the \f\ and the production threshold of two neutral 
kaons (995 MeV). The signature of this decay is provided by the 
presence of either 2 \ks\ or 2 \kl\ and a low energy photon. 
In the reported analysis, only the \ks\ks\ component has been used, 
looking for double \ks\to\pip\pim decay vertex. 

\noindent Theory predictions on the BR(\f\to\kk$\gamma$) spread over 
several orders of magnitude. The latest evaluations 
essentially concentrate in the region of $10^{-8}$ 
(Fig.~\ref{fig:all}-bottom right). Several of them are ruled 
out by our result.

\noindent At the end of the analysis we observe 1 candidate 
with 0 expected background. The upper limit 
of the signal expectation is evaluated to be 
$S_{UL} = 3.9~@~90\%~C.L.$ which correspond to an upper limit on 
the BR of:
\vspace{-0.2cm}
\be
  BR(\f\to\kk\gamma) <  \frac{2 \cdot S_{UL}}{\lint \cdot \sigma_{\phi} \cdot
    BR^2(\ks\to\pip\pim)\cdot \varepsilon}
  = 1.8\cdot10^{-8},
  \vspace{-0.2cm}
\ee
where the factor 2 accounts for the multiplicity in the initial state, 
$\varepsilon$ is our signal efficiency, \lint\ is the integrated luminosity 
and $\sigma_{\phi}$ is the \f\ production cross section. 

\vspace{-0.5cm}
\section{\boldmath $\eta$ physics}
\vspace{-0.2cm}
%
%

In this paper, we report the best measurement of the $\eta$ 
mass to date using the  $\phi \to \eta \gamma$ decay.
This decay chain, assuming the $\phi$ meson at rest,  
is a source of monochromatic $\eta$-mesons of 363 MeV/c, 
recoiling against a photon of the same momentum. Detection 
of such a photon tags the presence of the $\eta$-meson. 
Photons from $\eta \to \gamma \gamma$  cover a continuum 
flat  spectrum between $147 < E_{\gamma} < 510$ MeV in the 
laboratory reference frame. 
The accuracy of the kinematic reconstruction of the event is 
due to the precise measurement of the photon emission
angles.
Due to the stability of the calibration for the 
detector and the very large sample
of $\eta$-mesons collected, we have been able to obtain a 
very precise measurement of the $\eta$-mass~\cite{nota}.

\noindent The systematic uncertainties are from to detector 
response and alignment, event selection cuts, kinematic fit 
and beam energy calibration. We obtain the most
accurate measurement so far:
\vspace{-0.1cm}
\be
m_{\eta} = (547.873 \; \pm 0.007_{\rm stat} \; \pm 0.031_{\rm syst}) \quad
\mbox{MeV},
\vspace{-0.1cm}
\ee

%
%

The decay of the isoscalar $\eta$ into three pions occurs through 
isospin violation and thus is sensitive to the up-down
quark mass difference.
Neglecting electromagnetic corrections, the decay amplitude is 
parametrised in terms of kinetic energy of the three 
pions\cite{etappp_arch}. Following the conventional notation, 
the decay amplitude is expanded around the center of the Dalitz plot
(X=Y=0) in powers of X and Y as:
$\vert A(X,Y)\vert^{2} \simeq 1 + aY + bY^{2} + c X + dX^{2} + eXY + fY^3 + ..$
\noindent
and the parameters ($a,b,c,d,e,f,..$ ) of the expansion are fitted to
the experimental data. The KLOE results in 
$10^{-3}$ units are:
\vspace{-0.1cm}
\be
  a = -1090\pm5^{+8}_{-19} \qquad  b=124\pm6\pm10 \qquad
  d=57\pm6^{+7}_{-16} \qquad f=140\pm10\pm2
\vspace{-0.1cm}
\ee

The \emph{f} parameter was totally unexpected, and the value of the \emph{b}
is lower of what expected in the current algebra assumption ($b=a^2/4$). 
In addiction, this study allows us to set the most accurate limit on the
Dalitz plot asymmetries which are sensitive to charge conjugation violation.
 
%
%
The $\eta\to 3\pio$ decay is a major decay mode of the $\eta$ despite 
the fact that is a G--parity forbidden transition. 
This decay is due almost exclusively to the isospin breaking part of QCD.
For the decays into three identical particles, it is possible to use a
symmetrical Dalitz plot where the event density is described as a
function of a single variable z:  
$\vert A_{\eta\to 3\pio}\left(z\right) \vert^{2} \sim 1 + 2\alpha z$.

\noindent $\alpha$ represents the difference from pure phase space.
The lowest order  predictions of Chiral Perturbation Theory quote a
zero value for $\alpha$.
KLOE results has been obtained using the Dalitz density
normalized with the MC expectation for pure phase space decay
(Fig.~\ref{fig:all}-top right).
The preliminary result~\cite{eta3p0_arch} is:
\vspace{-0.1cm}
\be
  \alpha = -0.027 \; \pm 0.004_{\rm stat} \; {^{+0.004}_{-0.006}}_{\rm syst}
\vspace{-0.1cm}
\ee

%
%

The study of the $\etappee$ decay allows to probe the internal 
structure of the $\eta$ meson~\cite{Landsberg} and could be used 
to compare different theoretical  
predictions~\cite{etappeepred}.
Moreover, it would be possible to study CP violation not 
predicted by the Standard Model~\cite{Gao:2002gq}.
This can be experimentally tested by measuring the angular asymmetry 
between pions and electrons decay planes.
The preliminary measurement presented here is based on 600 pb$^{-1}$,
and is the first observation of such a process.
The number of selected signal events ($N_{events} = 733 \pm 62$) 
has been determined using MC shapes for signal and backgrounds
(Fig.~\ref{fig:all}-top left), we obtain:
\vspace{-0.1cm}
\be
BR(\etappee) = 
     ( 24 \; \pm 2_{\rm stat} \; \pm 4_{\rm syst} ) \times 10^{-5}.
\vspace{-0.1cm}
\ee
The selection efficiency of our signal, $\epsilon$, is evaluated 
by MC to be $\epsilon = 0.1175(5)$. 

%
%

KLOE has studied the $\eta/\eta'$ mixing~\cite{rphi_glue} using the 
measured $R_{\f}=BR(\f\to\eta'\gamma)/BR(\f\to\eta\gamma)$ following
the prescription of Bramon {\it et al.}~\cite{Bramon}. 
In our analysis we allow for a gluonium content in the $\eta'$ meson 
wave function even if the model parameters used in 
the analysis was calculated with the opposite assumption 
by Bramon (\emph{e.g.} no gluonium content). 
Further checks has been performed to validate our previous results.
Up to now the fraction of gluonium in the wave 
function is stable whit respect to the variation of the model parameters. 

\vspace{-0.5cm}
\section{Continuum process}\label{sec:cont}
\vspace{-0.2cm}

Using $\sim$ 600 pb$^{-1}$ collected at center of mass 
energies between 1000 and 1030 MeV, the cross sections
of \eeto\wpo\ in two different final states have been 
studied~\cite{WP0arch}: \wpc\ and \wpn.

\begin{figure}[!ht]
  \begin{center}
    \epsfig{figure=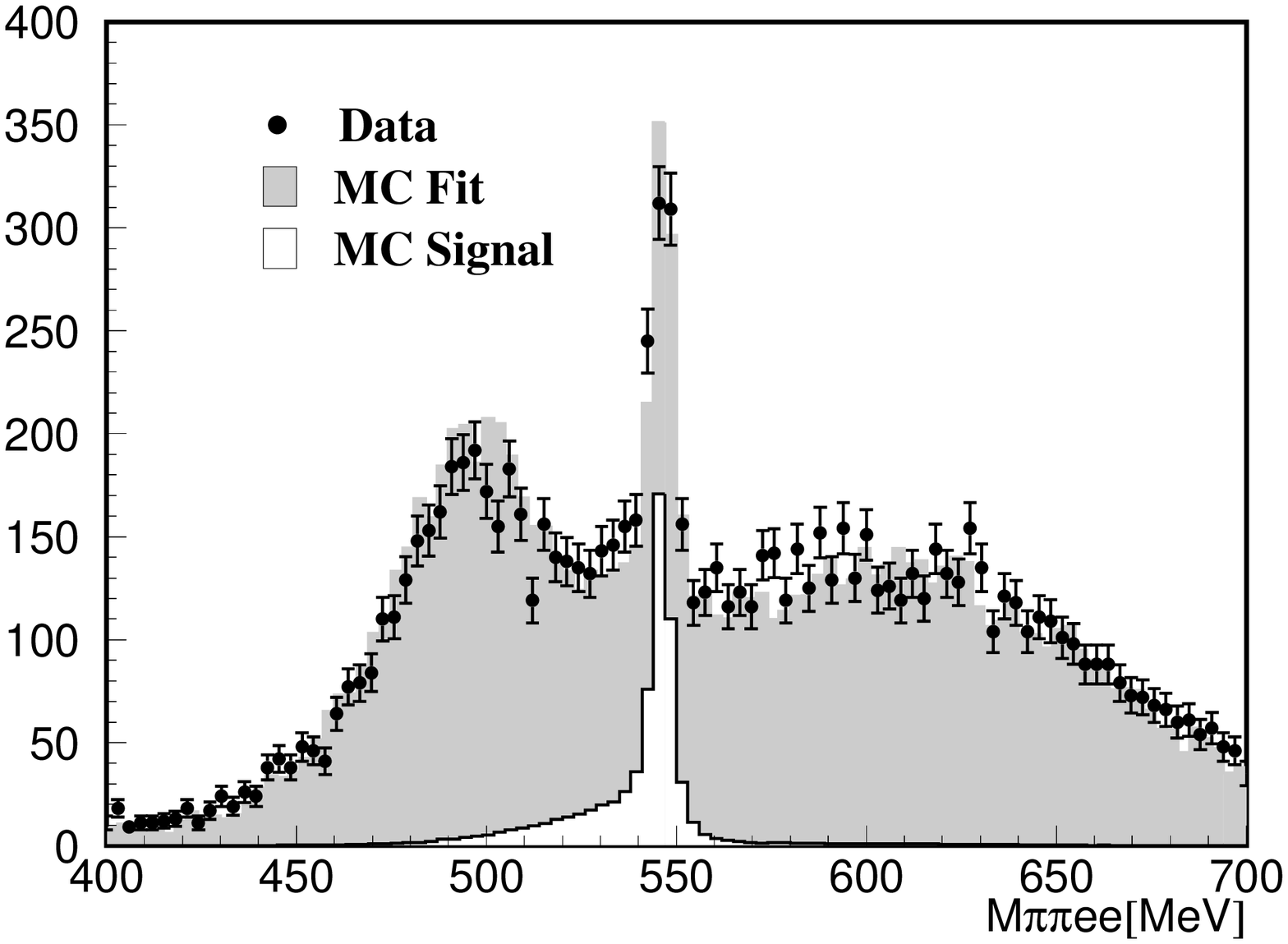, width=0.42\textwidth}
    \epsfig{figure=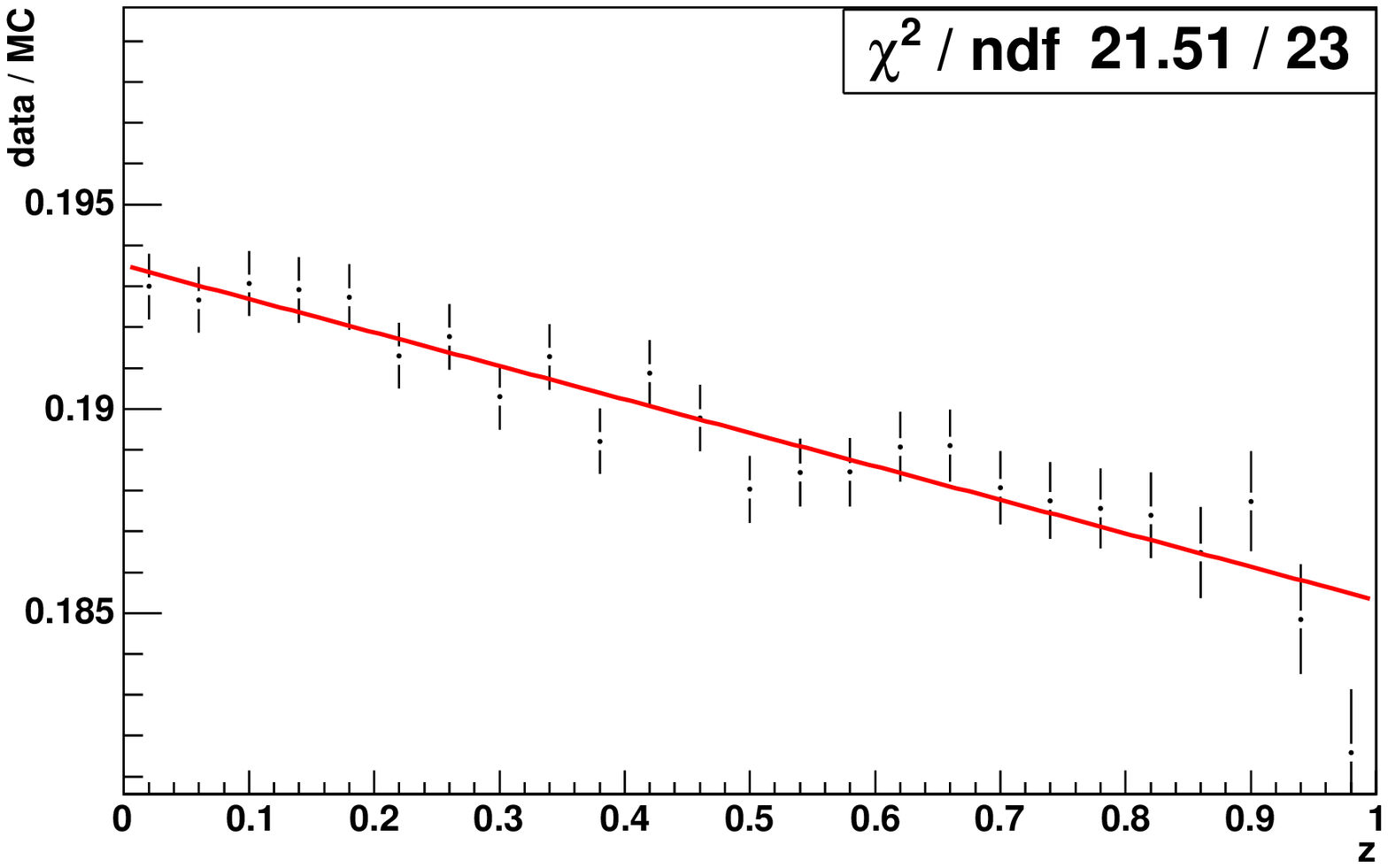, width=0.54\textwidth}
    \epsfig{figure=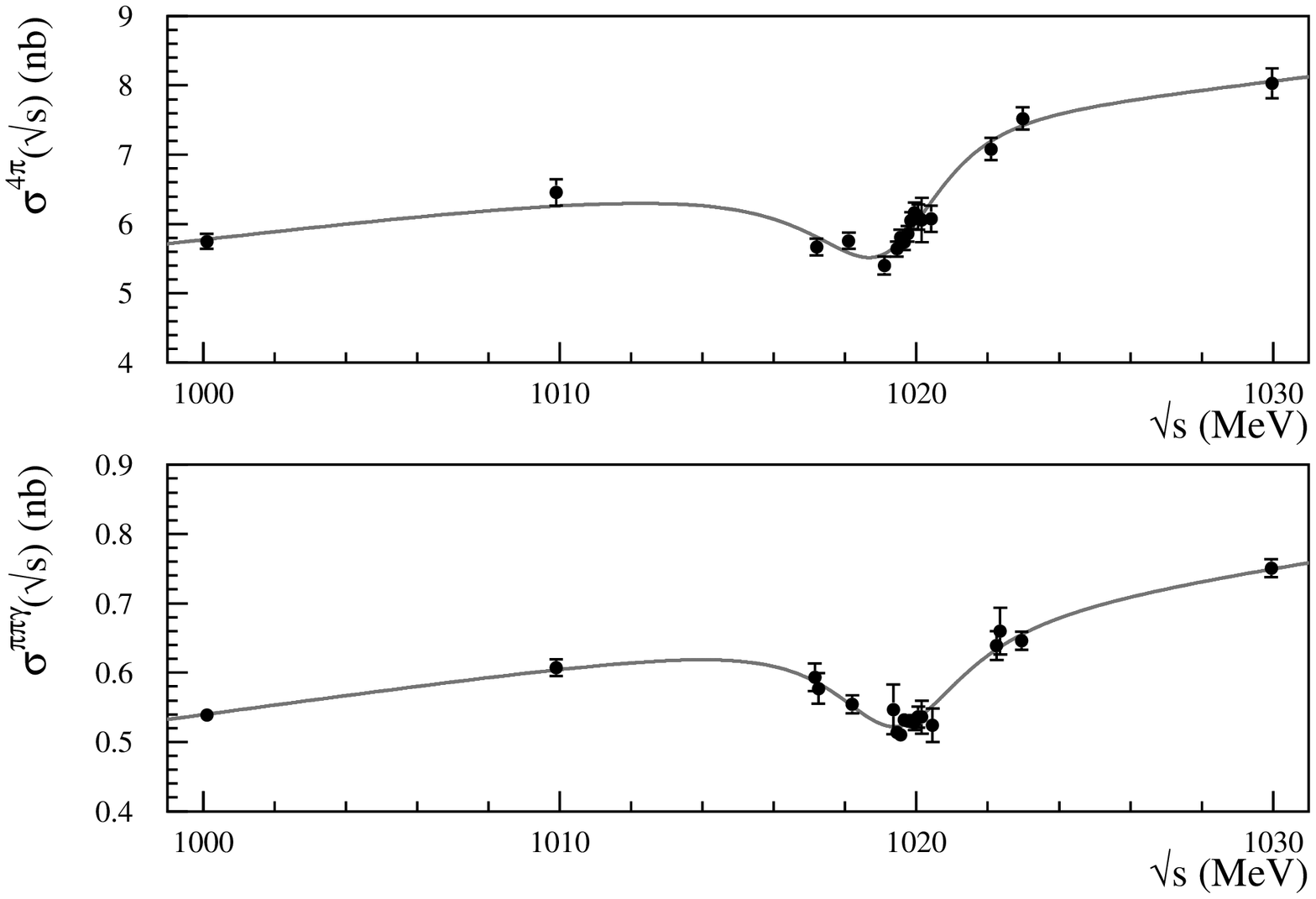, width=0.55\textwidth}
    \epsfig{figure=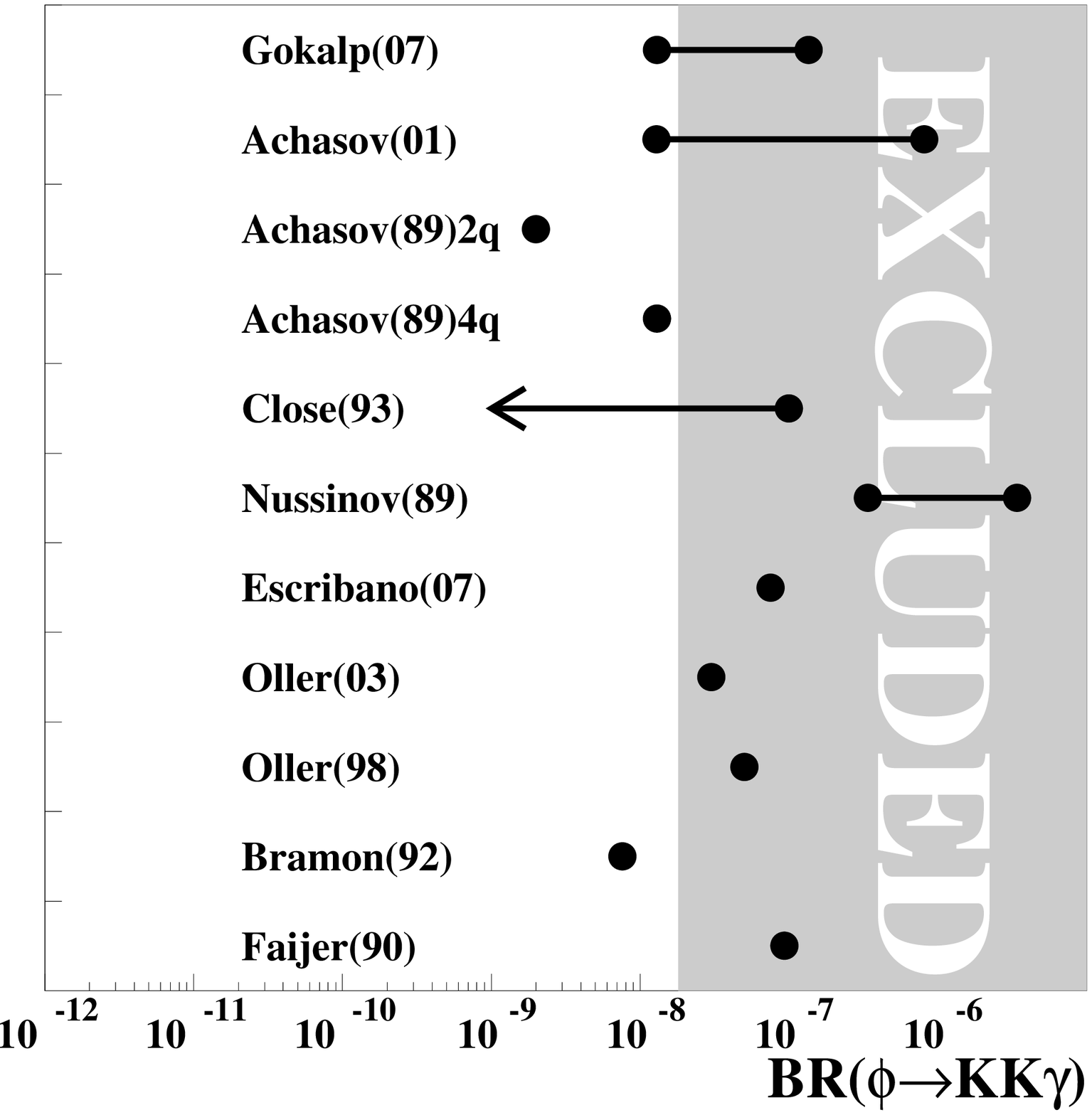, width=0.36\textwidth}
  \end{center}
  \caption{From top-left to bottom-right: $\eta$ mass fit in $\etappee$ 
    analysis, slope $\alpha$ for $\eta\to3\pio$, visible cross section 
    fit for the \eeto\wpc(top) and \eeto\wpn(bottom) channels, UL on 
    BR(\f\to\kk$\gamma$) exclusion plot.}
  \label{fig:all}
\end{figure}

\noindent In this energy region, the production 
cross section for both final states is largely dominated by the 
non-resonant processes $\eeto \rho/\rho' \to \omega \pio$. 
However, in a region closer to $M_\phi$, a contribution 
from the decay \f\to\wpo\ could be observed as an interference 
with the non-resonant processes.The cross section as a function of \sqrts\
can be parametrized in the form~\cite{phiwp2}:
\vspace{-0.1cm}
\be \label{eq:cont:xsec}
\sigma(\sqrts) = \sigma_{0}(\sqrts)
\cdot\left|1-Z\,(m_{\phi}\Gamma_{\phi}\Pi_{\phi})\right|^2
\qquad
\Pi_{\phi} =  (m_{\phi}^2 - \sqrts^2 - i\sqrts\Gamma_{\phi})^{-1} 
\vspace{-0.1cm}
\ee
where $Z$ is the complex interference parameter 
(i.e. the ratio between the $\phi$ decay amplitude and the non 
resonant processes), $\sigma_{0}(\sqrts)$ is the non-resonant
cross section while $m_{\phi}$, $\Gamma_{\phi}$ and $D_{\phi}$
are the mass, the width and the inverse propagator of the 
$\phi$ meson respectively. For $\sigma_{0}(\sqrts)$ a linear 
approximation has been used. The measured visible cross section
for both final state are fitted with Eq.~\ref{eq:cont:xsec}
convoluted with radiator function~\cite{RAD} 
(Fig.~\ref{fig:all}-bottom left). 

\noindent Using the cross section, the partial decay width ratio has 
been obtained: 
$\Gamma(\omega\to\pio\gamma)/\Gamma(\omega\to\pip\pim\pio)=0.093\pm 0.002$

\noindent Since these two final states correspond to the $98\%$ of 
the $\omega$ decay width, we use the ratio together with the 
unitarity~\cite{PDG07} to extract the main $\omega$ 
branching fractions:
\vspace{-0.1cm}
\be \label{eq:wbr}
  BR(\omega\to\pip\pim\pio)  =  (89.94\pm 0.23)\%  \quad 
  BR(\omega\to\pio\gamma)    =  (8.40\pm 0.19)\% 
\vspace{-0.1cm}
\ee

\noindent The measured parameters for \wpo\ cross section
in \wpc\ final state
and the peak value of the bare production cross 
section for the \f\ resonance~\cite{GeeKLOE}
are related to the BR(\f\to\wpo) through the relation:
\vspace{-0.1cm}
\be \label{eq:phiwpo}
  BR(\phi\to\wpo) = \sigma_0(m_{\f})|Z|^2/\sigma_{\f}
  = (5.63\pm 0.70)\times 10^{-5}
\ee

\noindent
Results \ref{eq:wbr} and \ref{eq:phiwpo} improves by a factor 
two the accuracy whit respect to the previous determination.
%

\end{document}